
\magnification=1200
\baselineskip 19pt
\parskip=6pt
\tolerance=10000 \hyphenpenalty10000 \exhyphenpenalty10000
\overfullrule=0pt
\def\cl{\centerline}
\def\page{\vfill\eject}

\def\etal{{\it et al.\ }}
\def\rms{{\it rms\ }}
\def\ltsima{$\; \buildrel < \over \sim \;$}
\def\lsim{\lower.5ex\hbox{\ltsima}}
\def\gtsima{$\; \buildrel > \over \sim \;$}
\def\gsim{\lower.5ex\hbox{\gtsima}}

\def\kms{\ {\rm km\,s^{-1}}}
\def\hmpc{\ {\rm h^{-1}Mpc}}

\def\dd{d}
\def\ln{{\rm ln}}
\def\pa{\partial}
\def\la{\langle}
\def\ra{\rangle}
\def\ti{\tilde}
\def\pot{$\scriptstyle POTENT$}
\def\iras{$\scriptstyle IRAS$}
%
\def\pmb#1{\setbox0=\hbox{#1}%
 \kern-.025em\copy0\kern-\wd0
 \kern.05em\copy0\kern-\wd0
 \kern-.025em\raise.0433em\box0}
\def\vv{\pmb{$v$}}
\def\vq{\pmb{$q$}}
\def\vx{\pmb{$x$}}

\def\vpsi{\pmb{$\psi$}}
\def\vnabla{\pmb{$\nabla$}}

\def\divv{\vnabla\cdot\vv}

\def\lam{\lambda}
%
\def\Ps{{\cal P}}
\def\Qs{{\cal Q}}
\def\rhos{\varrho}
\def\vs{\vartheta}
\def\vvs{\pmb{$\vartheta$}}
\def\Vs{{\cal V}}
%
\
\vbox{\vskip 3truecm}
\cl{\bf EVOLUTION OF ONE-POINT DISTRIBUTIONS}
\cl{\bf FROM GAUSSIAN INITIAL FLUCTUATIONS}

\bigskip\bigskip\bigskip
\cl{Lev Kofman,$^{1,2,3}$
Edmund Bertschinger,$^{4}$
James M. Gelb,$^{4,5}$}
\cl{Adi Nusser,$^{6,7}$
and Avishai Dekel,$^{6,8}$}

\bigskip\bigskip\bigskip\bigskip

\centerline{Accepted:{\it The Astrophysical Journal} 1994,{\bf 420},
 January 1 }

\noindent\vfootnote{$^1$}{CIAR Cosmology program and CITA, University of
 Toronto, Toronto, ON M5S 1A7,Canada}
\noindent\vfootnote{$^2$}{On leave of absence from Tartu Astrophysical
 Observatory, Estonia EE-2444}
\noindent\vfootnote{$^3$}{Institute for Astronomy, University of Hawaii,
2680 Woodlawn Dr., Honolulu, HI 96822}
\noindent\vfootnote{$^4$}{Department of Physics, Massachusetts Institute of
Technology, Cambridge, MA 02139}
\noindent\vfootnote{$^5$}{Fermilab, MS 209, P.O. Box 500, Batavia, IL 60510}
\noindent\vfootnote{$^6$}{Racah Institute of Physics, The Hebrew University of
Jerusalem, 91904, Israel.}
\noindent\vfootnote{$^7$}{Department of Astronomy,
University of California, Berkeley, CA 94720}
\noindent\vfootnote{$^8$}{Institut d'Astrophysique and Observatoire de Paris,
France}

\page
\cl{\bf ABSTRACT}

 We study the quasilinear evolution of the one-point probability
density functions (PDFs) of the smoothed density and velocity
fields in a cosmological
gravitating system beginning with Gaussian initial fluctuations.
Our analytic results are based on the Zel'dovich approximation and
laminar flow.
A numerical analysis extends the results into the multistreaming
regime using the smoothed fields of a CDM N-body simulation.
We find
that the PDF of velocity, both Lagrangian and Eulerian, remains Gaussian
under the laminar Zel'dovich approximation,
and it is almost indistinguishable from Gaussian in the simulations.
The PDF of mass density deviates from a normal distribution
early in the quasilinear regime and it develops a shape remarkably
similar to a lognormal distribution with
one parameter, the \rms density fluctuation $\sigma$.
Applying these results to currently available data we find that the PDFs
of the velocity and density fields, as recovered by the \pot\ procedure
from observed velocities
assuming $\Omega=1$,
or as deduced from a redshift survey of \iras\ galaxies
assuming that galaxies trace mass,
are consistent with Gaussian initial fluctuations.

\bigskip
{\it Subject headings:}\ cosmology --- dark matter --- galaxies:
clustering --- galaxies: formation

\page
\cl{\bf 1. INTRODUCTION}


The ``standard" model for the formation of large-scale structure
is based on gravitational instability of small initial fluctuations
in the energy density. These are assumed to originate
from quantum fluctuations that were stretched to
large comoving scales during the inflation phase (see Efstathiou 1990
for a review).
The fluctuations are assumed to be a random field, i.e.
a set of random variables, one for each point in space,
which is fully specified by the $m$-point joint probability density
functions (hereafter PDFs; cf. Monin and Yaglom 1971).
The time evolution of the PDFs for $m \geq 2$
may be sensitive to the nature of the dark matter
(e.g. being baryonic or nonbaryonic, hot or cold;
cf. Trimble 1987 for a review).
For example, the effect of the dark matter on the two-point correlation
function (or its Fourier transform, the power spectrum),
which is a moment of the $m=2$ PDF, is well known.
But the {\it one-point}
PDF is not explicitly sensitive to the nature of the dark
matter, at least in the linear regime.
This makes it a useful statistic for relating the present
fluctuations to the initial conditions
independently of the nature of the dark matter.

The natural choice for the density field is here,
as in many other physical systems,
a Gaussian random field, where the one-point probability distribution of
the density fluctuation field, $\delta\equiv\delta\rho/\rho$, is normal
with zero mean:
$$
P(\delta)={1 \over (2\pi\sigma^2)^{1/2}}\ e^{-\delta^2/2\sigma^2} \ ,
\eqno (1)
$$
and the joint probabilities are multivariate normal distributions.
(Note that we use $P$ to denote the probability density, or frequency
function, and not the cumulative probability.)
In the Gaussian case all the
moments
 are determined by the
variance $\sigma$
and the Fourier components of the density field have random phases.
In the linear regime, the density fluctuations and the three components
of the peculiar velocity field, $\vv$, are related linearly:
$\delta \propto -\divv$, so each of the velocity components
is a Gaussian random field too.
The one-point PDFs of density and the three velocity components
are all independent.

However, observed correlations in the
distribution of galaxies and clusters on scales $\geq 20\hmpc$
(e.g. Maddox \etal 1990; Efstathiou \etal 1990; Bahcall 1988;
Olivier \etal 1993),
and the coherence of their velocities on large scales
(e.g., Lynden-Bell \etal 1988),
have motivated a consideration of non-Gaussian initial
fluctuations. This is because these observed correlations are in excess
of the predictions of the ``standard" CDM model
when normalized to fit the distribution of galaxies on smaller scales.
This model assumes Gaussian, adiabatic
initial fluctuations with a scale-free spectrum and cold dark matter
dominating an $\Omega=1$, $\Lambda=0$ Friedmann universe.

 Theoretically, it has been shown that the general inflation picture
still
permits a wide variety of non-Gaussian fluctuations within the
standard gravitational instability theory (e.g.  Linde 1990;
Kofman 1991a).
Non-Gaussian perturbations also arise in other scenarios, e.g. where the
fluctuations originate from topological defects, such as cosmic strings
(see Bertschinger 1989 for a review) or textures (Turok 1991),
or from non-gravitational cosmic explosions
(see Ostriker 1988 for a review).  The statistical nature of the
initial fluctuations
is therefore a basic distinguishing feature between major competing
theories.

Several recent observations of the large-scale fields
allow determinations of the PDFs of
density and velocity based on their spatial distributions in increasingly
large volumes.
Are these fields consistent with Gaussian initial fluctuations?
Can they be used to reject this or other hypotheses?
In order to be able to answer these questions we first need to study
how the PDFs evolve in time under gravity.  During linear evolution,
when all Fourier components evolve at the same rate, the
PDFs do not
change form.  However, nonlinear evolution can introduce strong
non-Gaussian features.

Weinberg and Cole
(1992) have compared the effects of
initial non-Gaussianity with nonlinear evolution
from Gaussian initial conditions
using a series
of N-body simulations.  They found that, while nonlinear evolution
produces non-Gaussian density distributions
which may smear out the initial conditions,
some features of the initial conditions are preserved, enabling one to
use the present PDF for distinguishing
certain models of non-Gaussian initial conditions from Gaussian ones.

In this paper we
study the mildly nonlinear evolution of the PDFs
from Gaussian initial conditions.
In \S 2 we present formal expressions for the weakly nonlinear one-point
density and velocity PDFs in terms of the initial PDFs
for general initial conditions.
We evaluate these expressions for an initial Gaussian random field
and then approximate the evolution
using the Zel'dovich formalism in the limit of laminar flow.
Then, in \S 3, we use a ``standard" CDM high-resolution
N-body simulation to test
the approximation and extend the results into the multistreaming regime.
Our results are applied in \S 4 to two derivations of the smoothed
velocity and density fields in our cosmological neighborhood: the \pot\
analysis of the observed radial peculiar velocities of galaxies
(Dekel \etal 1990; Bertschinger \etal 1990) and the fluctuation fields
deduced from a redshift survey of \iras\ galaxies (Strauss \etal 1992).
Our analysis and results are discussed in \S 5.

\bigskip
\cl{\bf 2. PDF EVOLUTION IN THE LAMINAR ZEL'DOVICH APPROXIMATION}
\smallskip

\cl{\bf 2.1. Lagrangian vs. Eulerian PDFs}

To avoid confusion, we introduce the basic concepts and our notations
methodically and in detail.
Consider a large comoving volume $V$ ($\rightarrow \infty$) in the space of
comoving positions $\vx$, which contains a total mass $M$.
Assume that at time $t=0$ the mass is
distributed uniformly in space among particles of identical,
infinitesimal masses $\dd m$ ($\rightarrow 0$), at initial (Lagrangian)
comoving positions $\vq$. Each particle is identified from then on by
its Lagrangian position $\vq$. Let
the Eulerian position of particle $\vq$ at time $t$ be $\vx(\vq,t)$.
If the mapping from $\vq$ to $\vx$ is one-to-one, we call the flow
laminar or single-stream.  If it is many-to-one, i.e., if more than
one $\vq$ arrives at the same $\vx$ at a fixed time $t$, we call
the flow nonlaminar or multistream.

Let $\rho(\vx,t)$ be the mass density at position
$\vx$ at $t$.  We may treat $\rho$ as a random field in Eulerian space:
$\rho$ is drawn at random from a probability density $P(\rho)$
at each position.
We can imagine an ensemble of density fields where, at each
position, $P(\rho)\dd \rho$ is the probability that $\rho$ is in the
range $(\rho,\rho+\dd\rho)$.

Define $\rho(\vq,t)$ over Lagrangian space to be $\rho[\vx(\vq,t),t]$,
the density in the $\vx$ position of particle $\vq$ at time $t$.
(Note that in the multistream case, more than one Lagrangian point
$\vq$ may correspond to the same density $\rho$.)
This is a random field over Lagrangian space:
$\rho$ is drawn at random from a probability density $Q(\rho)$
for any randomly chosen particle.
We again have in mind an ensemble
of realizations where, for each particle, $Q(\rho)\dd\rho$ is the
probability that the particle resides in a region of density in the range
$(\rho,\rho+\dd\rho)$.
The difference between $P$ and $Q$ is that the probability measure is based
on volume for $P$ and on mass (i.e., Lagrangian volume) for $Q$.

The PDFs for the one-dimensional components of the velocity, $P(v)$ and
$Q(v)$, are defined analogously.  The distributions of $\vv$ are assumed
to be isotropic, i.e., the marginal distributions of each component are
identical,  $P(v_x)=P(v_y)=P(v_z)$.

A general relation between the Lagrangian and Eulerian PDFs of density,
which will be useful later, is
$$
P(\rho) = {\bar \rho \over \rho}\ Q(\rho) \ , \eqno (2)
$$
where $\bar\rho\equiv\int\rho P(\rho) \dd \rho$ is the mean density.
One simple way to prove this is by using the
alternative, spatial interpretation of the distributions.
Assuming ergodicity in Lagrangian space [that $Q(\rho)$ is independent of
$\vq$], one can replace the distribution over the
ensemble of random realizations at a given $\vq$ with the distribution
over Lagrangian space in one realization, so $Q(\rho)\dd \rho$
is also the fraction of mass
which resides in regions where $\rho$ is in the specified range.
The total mass with that property is then $M Q(\rho)\dd \rho$.
Assuming ergodicity in Eulerian space [that $P(\rho)$ is independent of
$\vx$, which, by the way, follows from the ergodicity in Lagrangian space
when the mapping between the spaces is one-to-one and onto],
one can replace the ensemble distribution
with the spatial distribution such that $P(\rho)\dd \rho$
can also be interpreted as the fraction of
volume in which $\rho$ is in the specified range.
The total volume with that property is $V P(\rho) \dd\rho$.

Using this interpretation involving the total mass of the particles that
reside in regions of a given
density and the corresponding volume occupied by this mass,
it is clear that
$M Q(\rho) \dd \rho = \rho\ V P(\rho)\dd\rho$, which implies Eq. (2)
with $\bar \rho=M/V$.

\medskip
\cl{\bf 2.2. On Multistreaming}

Our aim in this paper is to calculate the mildly-nonlinear PDFs of
density and velocity at time $t$, given the distributions at an initial
time.
Consider first each particle on its own.
Imagine it to be a mass element $\dd m$ initially spread uniformly inside
an infinitesimal volume $\dd^3q$. Let its volume at time $t$ be
$\dd^3x_q$, where the explicit mapping from Lagrangian to Eulerian
space, $\vx(\vq,t)$, is provided by the dynamics.
Assuming mass conservation one can write for each
element separately $\rhos(\vq,t) \dd^3x_q =\bar\rho \dd^3q$, so
$$
\rhos(\vq,t) = \bar\rho \left\Vert {\partial \vx \over \partial \vq}
\right\Vert
^{-1} \ , \eqno (3)
$$
where the double vertical bars denote the Jacobian determinant.
We call this density of an individual mass element (or ``a stream"),
$\rhos$, the ``single-stream" density. It
is not necessarily the same as the true, ``total" density $\rho$
used in \S 2.1.
This is because the mapping $\vx(\vq)$ is not necessarily one-to-one.
The mass elements may overlap in certain places, i.e.
particles from different $\vq$'s can cross in the same $\vx$ at $t$.
The total density $\rho$ at position $\vx$ is the
sum of the contributions $\rhos_i$ from all the streams
that arrive at $\vx$ at that time.
For a laminar flow, $\rhos=\rho$.

The mapping also determines the comoving ``single-stream" velocity of
particle $\vq$ at time $t$: $\vvs(\vq,t)=\dd\vx(\vq,t)/\dd t$.
(Note that this coordinate velocity must be multiplied by the expansion
scale factor to give the proper peculiar velocity.)
The actual ``total" velocity $\vv$ at $\vx$ is the mass-weighted average
of the velocities of all the particles that cross there at $t$.  The
difference between $\vvs$ and $\vv$ is equivalent to that between
the velocity of a molecule and the fluid velocity of a gas of molecules.

Given the PDFs of the initial fluctuations, a specific mapping
$\vx(\vq,t)$ is used below (\S 2.3 and \S 2.4) to compute the Lagrangian
PDFs of the single-stream quantities $\rhos$ and $\vs$ at
time $t$, which we denote $\Qs(\rhos)$ and $\Qs(\vs)$ (where $\vs$ is
one component of $\vvs$).
In the case of {\it laminar flow},
the single-stream quantities are equal to the total ones so the
single-stream PDFs are equal to
the PDFs of the total quantities, $Q(\rho)$ and $Q(v)$. The desired
Eulerian $P(\rho)$ can then be extracted using Eq. (2) and $P(v)$ can be
extracted in a similar way.

The analytic results of this paper are limited to this
laminar-flow approximation.
It is, however, worthwhile to continue
the analysis a bit further in the general framework of multistreaming
in order to better understand the nature of the approximation
and to predict its range of validity.

Consider the general case in which multistreaming may be present.
Each Eulerian volume element $\dd^3x$ may then contain several
Lagrangian volume elements $\dd^3q_i$, with $i$ labeling the
streams.  The Eulerian and Lagrangian volumes are related by
$\dd^3x=J(\vq_i)\dd^3q_i$, where $J$ is the Jacobian determinant
appearing in equation (3).  We can define a total single-stream
volume by integrating over all mass, $\Vs\equiv\int J(\vq\,)\,
\dd^3q$.  In general, $\Vs>V$ because Eulerian volumes with multiple
streams are counted more than once.  The Eulerian single-stream PDF
$\Ps(\rhos)$ is now defined as the probability that a point
randomly selected from $\Vs$ has single-stream density in the
range $(\rhos,\rhos+\dd\rhos)$ (and similarly for $\vs$).
The difference between $P$ and $\Ps$ is that the probability measure
is based on $V$ in the first case and $\Vs$ in the second.
Using ergodicity, $\Ps(\rhos)$ is also the fraction of $\Vs$ that has
single-stream density in the appropriate range.  Using mass
conservation in analogy with Equation (2), one then obtains
$$
\Ps(\rhos) = {\bar\rhos \over \rhos} \Qs (\rhos) \ ,
\eqno (4)
$$
where $\bar\rhos\equiv \int\rhos \Ps(\rhos) \dd\rhos$
is the mean ``Eulerian" single-stream density,
not to be confused with the total
$\bar\rho=\int P(\rho)\dd\rho$ when multistreaming is important.

An indicator for the degree of multistreaming is given by
the mean number of streams at an Eulerian point,
$$
N_s \equiv {\Vs \over V}
= {\bar \rho \over \bar\rhos} \ .
\eqno (5)
$$
This parameter can be computed from $\Qs (\rhos)$ by
$$
N_s= \int {\bar\rho \over \rhos} \Qs (\rhos) \dd\rhos \ ,
\eqno (6)
$$
where the total $\bar\rho$ is determined by the initial conditions and
it changes in time only due to the expansion of the universe,
$\bar\rho \propto a^{-3}$ with $a(t)$ the expansion factor.
Equation (6) follows
from the normalization of $\Ps$ as a PDF, $\int \Ps(\rhos)\dd\rhos = 1$,
combined with relation (4)
with $\bar\rhos$ replaced by $\bar\rho/N_s$ according to the
definition (5).

Normally, $N_s=1$ at early times and $N_s$ remains close to unity as
long as the flow is mostly laminar.
If the mapping $\vx(\vq,t)$ is continuous, i.e., it is onto the
Eulerian space such that no empty regions are formed, $N_s \geq 1$.
Eventually it grows to $N_s \gg 1$ in the severe multistreaming regime.
Therefore, $N_s(t)$ can serve as an indicator for the deviation from
laminar flow.
We will estimate $N_s(t)$ analytically below using the
Zel'dovich approximation to provide the mapping $\vx(\vq,t)$.

Given the single-stream PDFs, what are the desired total PDFs in the
presence of multistreaming?  The result for the density may be written
$$
P(\rho) = {1\over N_s} \Ps \left({\rho\over N_s}\right) + \delta P(\rho)\ ,
\eqno (7)
$$
where $\delta P(\rho)$ has vanishing zeroth and first moments so that
$P(\rho)$ maintains the proper normalization and has the correct mean,
$\bar\rho$.  The correction $\delta P$ is induced by caustics of the
mapping $\vx(\vq)$.  An example of this behavior is provided by the
gravitational microlensing problem, whose mathematics corresponds
to the two dimensional Zel'dovich approximation extended beyond caustic
formation.
There the probability distribution function $P(A)$ of magnification $A$
(the analogue of density here) has the caustic-induced feature $\delta P(A)$
for large $A$ (Nityananda and Ostriker 1984).
The same effect is expected in three-dimensional dynamics, whether exact
or given by the Zel'dovich approximation.

As a rough approximation we may use equation (7) with $\delta P=0$,
which should be valid while $N_s(t)$ is close to unity.  However, this
is certainly
not an exact solution in general and it is not even guaranteed to be a
reasonable approximation.  For an exact solution one has to sum over a
combination of single-stream probabilities
under the constraint that the single-stream densities (or velocities) sum
up (or average) to the given total density (or velocity).
The true PDF can be schematically written as
$$
P(\rho) = \sum_{N=1}^\infty p(N)
\ti P\left(\sum_{i=1}^N \rhos_i=\rho\right) \ ,
\eqno (8)
$$
where $p(N)$ is the  probability that there are $N$ streams at a point
(with mean $N_s$) and $\ti P$ is the joint probability density for $N$ streams
at one point to sum up to a total density $\rho$, which is some function
of the single-stream PDFs under the constraint.
This nontrivial calculation is beyond the scope of this paper.

\medskip
\cl{\bf 2.3. Velocity PDF in the Zel'dovich Approximation}

In any isotropic cosmology, there is a statistical symmetry between
positive and negative peculiar velocities
(in particular $\langle \vv \rangle =0$),
which must persist in the nonlinear regime as well.
Therefore, nonlinear deviations of an initially Gaussian
velocity distribution are subtle;
they are reflected only in the fourth order
irreducible moment of the one-point distribution
--- the kurtosis ---
or in higher even moments, which characterize features such as
the sharpness of the peak and the extent of the tail of the distribution.
(Recall that we are considering the PDF for one of the velocity components,
e.g., $v_x$.)
A priori, even such deviations could be large, but we show below that
they are in fact remarkably small in quasilinear gravitating systems.

Let us assume that the mapping from Lagrangian space to Eulerian space
is given by the Zel'dovich approximation (Zel'dovich 1970),
$$
\vx(\vq,t) = \vq + D(t) \vpsi (\vq) \ , \eqno (9)
$$
where $D(t)$ is a universal function of time [$D(t)\propto a(t)$ in
a spatially flat, pressureless universe].
The comoving peculiar velocity of particle $\vq$ is then
$$
\vvs(\vq,t) \equiv \dot {\vx}_q = \dot D \vpsi \ , \eqno (10)
$$
and the single-stream density is given by (Eq. 3)
$$
\rhos(\vq,t) = {\bar \rho \over \left\Vert {\bf I} +
D {\partial \vpsi \over \partial \vq}
\right\Vert } \ , \eqno (11)
$$
where ${\bf I}$ is the unit tensor.  For the total density $\rho(\vx,t)$
one should sum equation (11) over all streams $\vq$ at $\vx$.

It is easy to see that the Lagrangian PDF of velocities, $\Qs(\vs)$,
is {\it time-invariant} under the Zel'dovich approximation aside
from a simple scaling of $\vs$ in time (Eq. 10).
For an initially Gaussian PDF we
thus have for any of the three components of the velocity, at any time $t$
as long as the Zel'dovich approximation is valid,
$$
\Qs(\vs)= {1\over [2\pi\sigma_\vs^2(t)]^{1/2} }
{\rm exp}\left[-{\vs^2 \over 2\sigma_\vs^2(t)}\right] \ ,
\quad
\sigma_\vs^2(t) = \dot D ^2 (t) \langle \vert \psi \vert ^2 \rangle \ .
\eqno (12)
$$
Note that the time-invariance of the form of the PDF holds regardless
of the form of the initial PDF.

But somewhat surprising is the fact that
for Gaussian initial fluctuations and
under the Zel'dovich approximation
the Eulerian PDF of single-stream velocities is in fact {\it equal} to
the corresponding Lagrangian PDF at all times,
$$
\Ps(\vs) = \Qs(\vs) \ , \eqno (13)
$$
so it is also time-invariant!

To prove this, return to the interpretation of ensemble distribution,
and let $\Qs(\vs,\rhos)$ be the bivariate Lagrangian probability density
for $\vs$ and $\rhos$. Then, by Eq (4), the corresponding joint
probability density in Eulerian space is
$\Ps(\vs,\rhos)=(\bar\rhos/\rhos)\Qs(\vs,\rhos)$,
so one can write the Eulerian PDF for velocities as
$$
\Ps(\vs) = \int {\bar\rhos\over\rhos} \Qs(\vs,\rhos) \dd \rhos \ .
\eqno (14)
$$

Eq. (13) follows if $\vs$ and $\rhos$ are statistically independent
so that the bivariate distribution factors into the product of univariate
distributions,
$$
\Qs(\vs,\rhos) = \Qs(\vs)\ \Qs(\rhos) \ . \eqno (15)
$$
In the Zel'dovich approximation (Eqs. 10 and 11) at a fixed time
the velocity of a particle is a
function of $\vpsi$ only, while the density at a particle
is a function of $(\partial \vpsi /\partial \vq)$ only.
Thus, if $\vpsi$ and its spatial derivatives are statistically
independent, then $\Qs(\vs,\rhos)$ can be separated as in Eq. (15).
This condition is automatically met for a Gaussian random field:
the covariance $\langle\psi_i\,\partial\psi_j/\partial q_k\rangle$
vanishes by isotropy when $\vpsi$ and $\partial\vpsi/\partial\vq$
are evaluated at the same position.  Vanishing covariance implies statistical
independence for a bivariate normal distribution.
(Equation 15 could be valid for other random fields as well but
this issue is beyond the scope of this paper; we therefore restrict the
following discussion to initially Gaussian fields.)

The integral over $\rhos$ (in Eq. 14) can now be performed for a fixed
$\vs$, yielding by Eqs (4) and the normalization of $\Ps$
$$
\Ps(\vs)
= \Qs(\vs)\ \int {\bar \rhos \over \rhos} \Qs(\rhos) \dd\rhos
= \Qs(\vs)\ \int \Ps(\rhos) \dd\rhos
= \Qs(\vs)\ ,
\eqno (16)
$$
proving Eq. (13).
Hence, since $\Qs(\vs)$ is time invariant under the Zel'dovich
approximation, $\Ps(\vs)$ is time invariant too.

Note that the above invariance is valid only for the one-point PDF.
The velocity field does not remain Gaussian, in the
sense that the distribution of velocities at different points
in space is not expected to remain a multivariate normal distribution.

We conclude that, as long as orbit crossing is negligible,
$P(v)=\Ps(\vs)$ is time invariant [aside from the trivial
time-dependence $\sigma_\vs(t)$] and should remain Gaussian.
In the case of multistreaming, $Q(v)$ and $P(v)$ are not necessarily
time invariant because, for example, the probability for a given
number of streams at a point (Eq. 8) varies with time.
Also, too far into the multistreaming regime the Zel'dovich
approximation itself breaks down.
We test below (\S 3) the validity of this result in the presence of
multistreaming using the smoothed velocity field in an N-body simulation.

\medskip
\cl{\bf 2.4. Density PDF in the Zel'dovich Approximation}

 The density PDF, contrary to the velocity PDF,
is strongly affected
by nonlinear effects.  For one thing, a symmetric distribution of small
density fluctuations (with $\langle \delta \rangle =0$) develops an
asymmetry because the positive fluctuations can grow to any large value
while the negative fluctuations are limited by definition to $\delta \geq
-1$ ($\rho \geq 0$).  This eventually results in a sharp drop in
$Q(\delta)$ and in $P(\delta)$ toward
$\delta = -1$.
Another important effect is due to the fact that positive
$\delta$'s are typically associated with collapse, and therefore tend
to occupy smaller volumes at later times, while negative $\delta$'s
typically occur in `voids' which expand in time.  This tends to shift
$P(\delta)$ from positive to negative values.  Finally, the formation of
pancakes with high densities produces an extended tail for $Q(\delta)$
and $P(\delta)$ at large $\delta$'s.

The key of this section is the computation of
the single-stream Lagrangian PDF $\Qs(\rhos,t)$ of an initially Gaussian
density field that has evolved under the Zel'dovich mapping.
The somewhat elaborate calculation can be summarized as follows
(cf. Kofman 1991b).
Define $\ti\rhos \equiv \rhos/\bar\rho$.
Based on continuity, we write the Zel'dovich density (11) as
$$
\ti\rhos(\vq,t) = \vert \nu_1(\vq,t) \vert ^{-1} \ ,
\quad \nu_1 \equiv (1-D\lambda_1)(1-D\lambda_2)(1-D\lambda_3) \ ,
\eqno (17)
$$
with $\lambda_i$ the eigenvalues of the deformation tensor
$-\pa \psi_i / \pa q_j$, provided as initial conditions.
The absolute value in the denominator allows one to continue using this
expression even after the particle has passed through a caustic (where
$\nu_1=0$).
For convenience we can write $\nu_1$ as a cubic
in $D(t)$,
$$
\nu_1 =  1-D\mu_1 +D^2\mu_2 -D^3\mu_3 \ ,
\eqno (18)
$$
$$
\mu_1\equiv \lambda_1+\lambda_2+\lambda_3,\quad
\mu_2\equiv\lambda_1\lambda_2+\lambda_1\lambda_3+\lambda_2\lambda_3,\quad
\mu_3\equiv\lambda_1\lambda_2\lambda_3 \ .
$$

The crucial input into the desired calculation is the joint probability
density for the
eigenvalues in
a Gaussian field, which has been computed by Doroshkevich (1970) to be:
$$
\Qs_\lam(\lam_1,\lam_2,\lam_3)=
{5^{5/2}\, 27 \over 8 \pi \sigma_{in}^6}\
(\lam_1-\lam_2)(\lam_1-\lam_3)(\lam_2-\lam_3)\
{\rm exp}\left[-{1\over\sigma_{in}^2}
(3\mu_1^2-7.5\mu_2)\right] \
, \eqno (19)
$$
with $\sigma_{in}$ equaling the variance of $\ti \rhos$
at some initial time when the field is Gaussian.
Equation (19) now allows us to determine the probability density
for $\ti\rhos$ as a function of the $\lam_i$'s through equation (17).

The calculation becomes easier if we replace the eigenvalues $\lam_i$ by
the more convenient variables
$$
\nu_1, \quad
\nu_2 \equiv D^2\mu_2 - D^3\mu_3, \quad
\nu_3 \equiv D^3\mu_3 \ ,
\eqno (20)
$$
the first of which is simply related to the desired $\ti\rhos$ (Eq.
17). Given this transformation,
the joint PDF of the new variables can be expressed in terms of the PDF
of the old variables:
$$
\Qs_\nu(\nu_1,\nu_2,\nu_3) =
\Qs_\lam(\lam_1,\lam_2,\lam_3) \,
\left\Vert {\pa (\nu_1,\nu_2,\nu_3)
\over \pa (\lam_1,\lam_2,\lam_3)} \right\Vert^{-1}\ .
\eqno (21)
$$
Then, the PDF of $\nu_1$ is given simply by double integration,
$$
\Qs_{\nu 1}(\nu_1) = \int \dd\nu_2 \int \dd\nu_3 \,
\Qs_\nu (\nu_1,\nu_2,\nu_3) \
\eqno (22)
$$
over the appropriate range in the $(\nu_1,\nu_2,\nu_3)$ space.
The desired PDF of relative density, $\ti\rhos =\vert \nu_1 \vert^{-1}$,
is then given by
$$
\Qs(\ti \rhos) = \ti\rhos^{-2}
     [\Qs_{\nu 1}(\ti\rhos^{-1} )
     +\Qs_{\nu 1}(-\ti\rhos^{-1}) ]\ .
\eqno (23)
$$

What is left is to express the $\lam_i$'s in term of the $\nu_i$'s
and the hard part is to obtain the limits of integration in the double
integral (22). We note when inverting the transformation (20)
that the $\lam_i$'s can be evaluated by solving
the cubic polynomial equation
$$
\lam^3
-D^{-1}(1-\nu_1+\nu_2)\lam^2
+D^{-2}(\nu_2+\nu_3)\lam
-D^{-3}\nu_3  =0 \ .
\eqno (24)
$$
The problem of defining the range of integration in (22)
is thus reduced to finding where all three roots
of Equation (23) are real. In practice we simply set $\Qs_\nu=0$
when the roots are not all real.

We finally obtain after some algebra (see Appendix A for details)
$$
\Qs(\ti\rhos,t) = {N  \over \ti\rhos^2 \sigma^4}
\int_{3 \tilde\rhos ^{-1/3}} ^\infty  \dd s\
e^{-{(s-3)^2 / 2 \sigma^2}}
\left( 1+ e^{-{6s/ \sigma^2}} \right)
\ \left( e^{-{\beta_1^2 / 2\sigma^2}}
             +e^{-{\beta_2^2 / 2\sigma^2}}
             -e^{-{\beta_3^2 / 2\sigma^2}}  \right)
\ , \eqno (25a)
$$
$$
\beta_n (s) \equiv s\,5^{1/2} \left( {1\over2}
+\cos\left[{2\over3}(n-1)\pi
+{1\over3} \arccos \left({54 \over \ti\rhos s^3}
-1 \right)\right]\right) ,
 \eqno (25b)
$$
where $\sigma(t)\equiv D(t)\sigma_{in}$ is the standard deviation of
$\ti\rhos =\rhos/\bar\rho$ at time
$t$ according to linear theory  [given $\sigma_{in}$ at $t_{in}$ and
$D(t)$ being the growing solution between $t_{in}$ and $t$].
The shape of $\Qs(\ti\rhos)$ depends on time only via the parameter
$\sigma$. The numerical factor is
$N=9 \cdot 5^{3/2}/4\pi \approx 8.007328$.
$\int \Qs(\ti\rhos)\dd\ti\rhos =1$. In the pure laminar regime $A=1$.
The complicated expression (25) for $\Qs(\ti\rho)$
indeed reduces to a simple Gaussian distribution when $\sigma \ll 1$
(see Apendix B for a proof).
Otherwise, the one-dimensional integral of equation (25) has to be
performed numerically for a given $\sigma$ and $\ti\rhos$.

The desired single-stream Eulerian distribution can now be evaluated by
Eqs. (4) and (5):
$$
\Ps(\ti\rhos)= {1 \over \ti\rhos N_s} \Qs(\rhos) \ .
\eqno (26)
$$
In the laminar regime equations (25) and (26) give
the Zel'dovich approximations for
$Q(\rho/\bar\rho)$ and $P(\rho/\bar\rho)$.

We can use the derived $\Qs(\ti\rhos)$ of the Zel'dovich approximation to
estimate
the mean number of streams at each position, $N_s(\sigma)$, using Eq.
(6),
$$
N_s(\sigma) =\int {1\over\ti\rhos} \Qs(\ti\rhos)\dd\ti\rhos \ .
\eqno (27)
$$
The resultant $N_s$ as a function of $\sigma$ is shown in Figure 1.
$N_s$, much like $A$,
remains flat at unity until $\sigma \sim 1$, and then,
when orbit crossing becomes severe and the Zel'dovich approximation
breaks down in certain places, it shoots off to large values.
This growth is supposed to be proportional to $D^2$ for large $\sigma$
for the following reason.
Replace the integration over $\Qs(\rhos)\dd\rhos$ in Eq. (27) by
$$
N_s = \int \ti\rhos^{-1}
\Qs_\lam(\lambda_1,\lambda_2,\lambda_3) \dd\lambda_1\dd\lambda_2\dd\lambda_3
\ . \eqno (28)
$$
Recall from Eq. (18) that
$\ti\rhos^{-1} = \vert 1 - \mu_1 D +  \mu_2 D^2 - \mu_3 D^3 \vert $.
A simple symmetry argument guarantees that
$Q(\lambda_1,\lambda_2,\lambda_3)$ is invariant under
$(\lambda_1,\lambda_2,\lambda_3) \rightarrow
(-\lambda_1,-\lambda_2,-\lambda_3)$,
so $\la \mu_1 \ra = \la \mu_3 \ra =0$.
The deviation from $N_s=1$ is therefore only due to the term which
grows in time $\propto D^2$.

As a bonus from this analysis we can gain, for example, some insight
into the formation of pancakes in the Zel'dovich approximation.
Simulations indicate that the process of pancaking is rather typical
(Melott and Shandarin  1989; Nusser and Dekel 1990; Kofman \etal 1992).
Assume that there is a universal density fall off from the nearest
caustic plane, $\rho(x)$.
Such a one-to-one correspondence between $x$ and $\rho$
implies $P(\rho)\ \dd \rho \propto \dd x$.
Since we find in Eq. (25) that for large densities
$P(\rho) \propto \rho^{-3}$, we
get by integration $x\propto \rho^{-2}$, i.e. $\rho \propto x^{-1/2}$.
This is indeed consistent with the density
profile of caustics in developed pancakes
(see Shandarin and Zel'dovich 1989 for a review). Kofman (1991b),
Coles and Jones (1991) and Coles and Frenk (1991) have also made
a similar point.

The mean number of streams at a point can tell us in turn about the rate of
pancake formation.
One can write $N_s=p(1)+3 p(3)+ ...$,
where $p(N)$ is the probability for $N$ streams at a point (as in Eq. 8),
since only an odd number of streams is possible.
The normalization, on the other hand, guaranties that $p(1) + p(3) + ...=1$.
If $N_s$ is not too big, we can ignore the occurrence of 5 streams or
more and estimate the probability for (three-stream) pancakes to occur
at any Eulerian point: $p(3) \approx (N_s -1)/2$.

\bigskip
\cl{\bf 3. EVOLUTION OF PDF'S IN AN N-BODY SIMULATION}

To test the nonlinear effects including the effects of multistreaming,
the PDFs have been computed in a cosmological N-body
simulation. We use a particle-mesh code (Bertschinger and Gelb 1991)
with $256^3$ grid cells and $128^3$ particles
in a periodic cubic box of comoving size $200\hmpc$.
The initial conditions for the simulation are a random Gaussian
realization of the ``standard" CDM spectrum (Davis \etal 1985),
assuming $\Omega=1$ and $h=0.5$. The spectrum is normalized such
that today, based on linear growth, $\sigma_8^2\equiv \langle \delta ^2
\rangle$ in spheres of radius $8\hmpc$ is unity.
The expansion factor at that time is set to $a=1$.
Figure 2 shows the particle distribution in the simulation at $a=1$
in an arbitrary slice of thickness $25\hmpc$.

Continuous density and velocity fields do not exist in an N-body
simulation, where instead one has a point process.  However, continuous
fields may be defined by replacing each particle with a smoothing kernel.
The statistical properties of the point process are then reflected
in those of the continuous density and velocity fields, as discussed
in detail by Scherrer \& Bertschinger (1991) and Scherrer (1992).

The desired density and velocity fields are evaluated on a $80^3$
grid of comoving spacing $2.5\hmpc$.
The first operations are on the grid-cell scale: a trilinear
interpolation (cloud-in-cell assignment)
of mass and momentum to a grid point from the particles in its
neighboring cells,
followed by small-scale Gaussian smoothing of the mass density and
momentum density with a Gaussian window of radius $2.5\hmpc$.
The velocity at a grid point is then defined as the ratio of momentum to
mass density there. The purpose of this small-scale smoothing is to
obtain meaningful velocity values in grid points that reside in a
neighborhood of empty cells so that they are not spuriously assigned zero
velocity.

The density and velocity fields are then smoothed further on
a larger scale using a spherical Gaussian window of
radius $R_s$. The purpose of this smoothing is to reduce the
effects of nonlinearities by diminishing the density contrasts.
A range of $R_s$ is considered in order to span a range of
nonlinear effects.  Note that smoothing after nonlinear evolution
does not fully remove nonlinear effects, although the longest
wavelengths are expected to evolve according to linear theory.
We will see,
however,
that on sufficiently large scales, smoothing restores
approximately the linear evolution of the large-scale initial conditions.

The smoothed velocity and density fields on the grid, at different times
and with different smoothing lengths, are used to
construct the PDFs.

Figure 3 shows the Eulerian velocity PDF in the most nonlinear case
studied with the simulation: $a=1$
and $R_s=6\hmpc$, with $\sigma_v=277 \kms$ (and $\sigma_\delta=0.55$).
The error bars are the standard deviation of the mean in eight octants of
side $100\hmpc$ each. These are only rough estimates of the
true statistical uncertainties because only eight subvolumes were used.
The PDFs at earlier times and larger smoothings are very similar and are
therefore not shown.
They are all very much Gaussian, as predicted by the laminar
Zel'dovich approximation.
The apparent excess in the positive tail beyond $3\sigma_v$ is
at least partly due to a random deviation in the initial
conditions due
to the limited volume of the box; a similar excess shows up at all times.
Thus, the N-body simulation confirms the Zel'dovich
prediction that the velocity PDF of initially Gaussian fluctuations
remains Gaussian in the quasilinear regime, and it extends this result
into the multistreaming regime.

Figure 4 shows the Eulerian density PDFs at $a=1$ for three different
smoothing lengths: $R_s=18, 12$ and $6\hmpc$, corresponding to
$\sigma=0.11, 0.26$ and $0.55$ respectively.
The symbols are the means from eight octants in the simulation and the
error bars are the corresponding standard deviations of the mean
in the eight octants.
A range of $\sigma$ values could similarily be spanned by a sequence of
time steps with a fixed smoothing length.
The dependence of the PDF on $\sigma$ is similar but not identical
(see the discussion of moments and Figure 5 below).
We show the smoothing sequence here because it is
what one can obtain observationally.

The dashed lines are based on the laminar Zel'dovich approximation
[$\Ps(\ti\rhos)$ from $\Qs(\ti\rhos)$ of Eq. 25 and 26] for the
corresponding $\sigma$ values.
The laminar Zel'dovich approximation indeed provides an excellent
approximation to the simulated PDF in the intermediate, slightly
nonlinear case with $\sigma=0.26$. At $\sigma=0.55$ the Zel'dovich
approximation is still a very good approximation out to
$\rho/\bar\rho=3$ ($\approx 5.5\sigma$), but it starts to overestimate
the positive tail beyond that. The Zel'dovich power-law tail,
$P\propto\rhos^{-3}$, reflects the collapse
of the highest peaks (in $\lambda_1$)
into caustics ---  caustics which
are smeared out by the smoothing applied to the N-body simulation.

This example shows how smoothing and nonlinear evolution do not always
commute.  If the initial density field is smoothed first, nonlinear
evolution produces caustics of high density.  However,
when an unsmoothed field is evolved and then smoothed,
the smoothing reduces the high densities.
Equation (25) is appropriate in the former case but not the latter.
Nonlinear evolution erases some memory of the initial conditions on
small scales but it also generates small-scale structure from the
collapse of long waves (Kofman \etal 1992).

The solid curves are lognormal distributions with the same $\sigma$,
$$
P(\ti\rho) = {1 \over (2\pi\sigma_l^2)^{1/2}}
{\rm exp}\left[{ ({\rm ln}\ti\rho - \mu_l)^2 \over \sigma_l^2} \right]
\cdot {1 \over \tilde\rho}\ ,
\eqno (29)
$$
where $\mu_l$ and $\sigma_l$ are the mean and standard deviation of
$\ln\ti\rho$. They are related to the corresponding moments of
$\ti\rho$, $\mu$ and $\sigma$, via
$$
\mu_l=\ln\mu-(1/2)\sigma_l^2 , \quad
\sigma_l^2=\ln(1+\sigma^2/\mu^2) \
\eqno (30)
$$
(and recall that in fact $\mu=1$ for $\ti\rho$).
As argued by Coles and Jones (1991) and noted earlier by
Hamilton (1985),
the lognormal distribution
turns out to be an excellent approximation to the actual density PDF.
The way it fits the simulation at $\sigma=0.55$ over the whole range
tested, out to $\sim 10\sigma$, is striking!

A more quantitative measure of the deviations of the PDFs from Gaussian
is provided by their third and forth irreducible
moments (Figure 5). Given a set of random measurement of the random
variable $x$, define as usual the mean,
$\mu\equiv \la x \ra$,
and the variance,
$\sigma^2\equiv \la (x-\mu)^2 \ra$.
Then the dimensionless skewness and kurtosis relative to the mean
are defined by
$$
s\equiv {\la (x - \mu)^3 \ra \over \sigma^3}
\quad {\rm and} \quad
k\equiv {\la (x - \mu)^4 \ra \over \sigma^4} -3 \ .
\eqno (31)
$$
As a reference, a Gaussian distribution has $s=k=0$.

Based on the Zel'dovich approximation, we expect that any deviation
from a normal shape are predominantly a function of $\sigma$.
We therefore plot in Figure 5 the skewness and kurtosis of the
PDFs as a function of $\sigma$.
Note though that a
range of $\sigma$ values could be obtained either by analyzing
the system at different times or by using a
variety of smoothing lengths.
Each panel shows three curves corresponding to three different
times ($a$), with the
Gaussian smoothing length ($R_s$ in comoving $\hmpc$) varying along each
curve.
The error bars mark the standard deviation of the mean for the moments as
evaluated in eight octants.

 Because of the expected symmetry between positive and negative
velocities, there is no surprise in the fact that the velocity skewness
is consistent with zero; any deviation must be a result of the limited
volume sampled.  But the fact that the velocity kurtosis remains constant
and consistent with Gaussian is a meaningful confirmation of the laminar
Zel'dovich approximation.  One can see that the apparent small deviation
from zero of both $s$ and $k$, which is probably associated with the
apparent tail in Figure 3, is indeed similar at all times, indicating
that it must be due to the finite volume sampled rather than nonlinear
evolution.

 While the moments of velocity remain constant over the whole range of
$\sigma_v$ tested, the moments of density gradually deviate from Gaussian
in the corresponding range of $\sigma_\delta$.  This deviation is
significant already at relatively low $\sigma_\delta$ values.

 We can also see from Figure 5 that the growth of $\sigma_\delta$ in
the simulations is very similar to the prediction of linear theory:
$\sigma_\delta \propto a$ to within 3\% in the tested range.  This is
saying that the faster nonlinear growth in high density regions is
roughly compensated for by the slower deepening of low-density regions
(limited by $\delta \geq -1$).  This allows one to use in the Zel'dovich
approximation (25) the true, observable $\sigma$ instead of the linear
$\sigma$ which we can't measure.

 The actual values of $a$ and $R_s$ that boil down to a given value of
$\sigma$ make some difference, which is significant (in view of the
errors) only for the density kurtosis.  This difference depends on the
power spectrum used.  Since observationally $\sigma$ can vary only due to
the smoothing used, we limited ourselves in Figures 3 and 4 to
``observing" the simulation only at one time while the smoothing
length is varied.  Assuming a universal galaxy biasing factor of unity
($b=\sigma_8^{-1}=1$), we used the time step in the simulation $a=1$.  If
the true biasing factor is different from unity, then a different time step in
the simulation should have been used to resemble the present universe
(e.g. $a=0.5$ for $b=2$).  Based on Figure 5, this would not affect much
the dependence of skewness on $\sigma$ but the density kurtosis would
behave somewhat differently.

 Second-order perturbation theory predicts that the ratio of density
skewness to its standard deviation is constant, $s/\sigma=34/7$ (Peebles 1980),
which is marked by the dotted line in the top-right panel of Figure 5.
We can see that this is a good approximation for $R_s \geq 12\hmpc$,
independently of how $\sigma$ was changed (by $a$ or by $R_s$).
At smaller smoothing lengths the N-body results tend towards $s/\sigma$
values in the range 4 to 3,
in general agreement with the various models discussed by Coles and
Frenk (1991). In particular, it has been predicted based on second order
theory that this ratio would depend on the
effective logarithmic slopes of the power spectrum at the smoothing scale,
$n$, roughly as $s/\sigma = 34/7 - (n+3)$
(Juszkiewicz, Bouchet and Colombi 1992,
but note they assumed top-hat smoothing).
In our case of a CDM spectrum and Gaussian smoothing lengths in the range
$6-21\hmpc$ the effective slope is in the vicinity of $n\approx -1$,
which indeed predicts $s/\sigma \approx 3$.

\bigskip
\cl{\bf 4. TENTATIVE COMPARISON WITH OBSERVATIONS}

Equipped with the above results concerning the PDFs of
initially Gaussian fluctuations for a given $\sigma$,
we now make first attempts to reconstruct the one-point
spatial distribution functions of the smoothed velocity and density
fields as estimated from galaxies in a finite volume around us.
The following comparison is tentative as it is based on limited
data in a relatively small volume.
The pilot data are provided by either the early \pot\
analysis of observed velocities or the analysis of the 1.9Jy
redshift survey of \iras\ galaxies.

 The \pot\ analysis used the observed radial peculiar velocities of
about 1000 galaxies in a sphere of radius $\sim 60\hmpc$ about the
Local Group. The observed radial velocities were first smoothed into
a radial velocity field
on a grid, in a way that minimizes the
effects of sparse sampling and measurement errors.  \pot\ then
imposes
the requirement of potential flow, $\vv = -\vnabla \phi$,
which is a natural outcome of gravitational instability, to
reconstruct the missing two components of the
velocity field (Bertschinger and Dekel 1989;
Dekel \etal 1990; Bertschinger \etal 1990).
Assuming that the galaxies are fair tracers
of the smoothed velocity field independent of their specific type,
the resultant velocity field is independent of galaxy ``biasing".
Finally, assuming a value for
$\Omega$, and using a quasilinear approximation for the
equations governing the evolution of fluctuations (Nusser \etal 1991),
\pot\ yielded the mass-density fluctuation field that has given
rise to the peculiar velocities.
The output is provided on a cubic grid of spacing $5\hmpc$ inside a
spherical volume of radius $60\hmpc$.
A Monte Carlo analysis of distance-measurement errors
provided estimates of the uncertainties of each field,
which enables us to use for the reconstruction of the PDFs
only points where the uncertainty is smaller than a certain
conservative limit, and to estimate the resultant uncertainties
in the PDFs.

Because of the limited volume analyzed,
and the zero-point uncertainty in the distance indicators,
the mean density and velocity
within this volume have finite values different from the universal zero
means: the mean density contrast is $\mu_\delta = 0.11$ and the mean
one-dimensional velocity is $\mu_v=223\kms$. We use the fact that if
$x$ is a Gaussian field then the conditional probability of $x$ in a
neighborhood where the local mean is given is also Gaussian
with a displaced mean and somewhat reduced dispersion (depending
on the two-point correlation function; cf. Dekel 1981, the appendix).
We therefore compute and plot the PDFs of
$(\delta-\mu_\delta)/\sigma_\delta$ and $(v-\mu_v)/\sigma_v$.

These PDFs of \pot\ velocities and densities, assuming $\Omega=1$,
are shown in Figure 6.
We use only points with Monte Carlo measurement errors in the
three-dimensional velocities and in the densities smaller than $300\kms$
and $0.3$ respectively.
The resultant effective volume corresponds to a sphere of radius
$37\hmpc$. The error bars are the standard deviation of the PDF in
30 Monte Carlo noise simulations of \pot.
Also marked are the derived first four moments.

Errors due to the limited volume sampled can be estimated using the CDM
N-body simulation, because the power-spectrum deduced from \pot\ is
not significantly different from the standard CDM spectrum
with $b=1$ (Kolatt, Seljak, Bertschinger and Dekel 1993).
In Figure 7 we show the mean PDF of eight
independent spheres of radius $50\hmpc$ from the simulation and the
associated standard deviations are marked by the error bars.
Also marked are the first four moments and their standard deviations.
The relative volume errors for \pot\ should be larger
by a factor which is roughly the square root of the volume ratio, i.e.
1.57.  For a conservative error estimate, the volume errors and the
measurement errors shown in figure 7 should be added in quadrature.

We see from figure 6 that the PDFs based on \pot\ are consistent with
Gaussian initial fluctuations. However, the errors are very big,
making this preliminary comparison only marginally interesting.
Things are expected to get better though because
the rapid progress in peculiar velocity surveys
allow a \pot\ reconstruction with smaller uncertainties
in a larger volume.

 The \iras\ analysis (Strauss \etal 1990; 1992; Yahil \etal 1990) translated
the redshift catalog of 1.9Jy \iras\
galaxies into a uniform galaxy-density map, whose first
approximation was given in redshift space.  The predicted peculiar
velocities, and the corresponding corrections to the galaxy
positions from redshift space to configuration space, were reconstructed
via a self-consistent iterative scheme using linear dynamical theory of
gravity with small-scale smoothing and quasilinear corrections (Nusser
\etal 1991), after assuming linear biasing between the density
fluctuations of galaxies and mass.
The resultant peculiar
velocity depends both on $\Omega$ and on the ``biasing" parameter $b$
between the density fluctuations of galaxies and mass, but
the obtained density field depends only weakly
on the assumed value of $\Omega$ through the
correction from redshifts to real positions.
This analysis provided estimates for the density and velocity
fields within a sphere of radius $80\hmpc$ about the local group.
In the present, preliminary application we take these fields as
provided to us by the authors of the \iras\ analysis; we do not make
an attempt to carefully estimate the errors involved in the sampling and
in the analysis beyond the volume errors.

A strong correlation is found between the \iras\ and \pot\ fields,
both featuring as extended structures
the Great Attractor, an adjacent large void, and the Pisces part of
the Perseus-Pisces supercluster.
The comparison yields $\Omega^{0.6}/b = 1.3 \pm 0.7$ (Dekel \etal 1993,
the error bar is 95\% confidence level).
This allows one to adopt the simple linear biasing assumption as a
working hypothesis and associate the PDFs derived from \iras\
galaxies with the desired PDFs of the underlying dynamical mass
distribution.

The distributions of the \iras\ densities and velocities, for
two different smoothing lengths, are shown in Figure 8.
The means are again removed and the deduced first four moments are marked.
The errors due to the finite volume should be about twice the error bars
shown in Figures 3 and 4 from the CDM N-body simulation, or about one
half of the errors in Figure 7. Within these uncertainties, the
\iras\ data are consistent with Gaussian initial fluctuations.
The ratio $s/\sigma$ is consistent with being a constant as a function
of smoothing scale, in the range $s/\sigma \approx 1.5 - 1.8$

Note that although \iras\  galaxies are underrepresented in rich
cluster cores, so that they are biased against high density, this
has little effect on $P(\rho)$ because clusters occupy a very small
fraction of the volume.  This is an advantage of the one-point
Eulerian density distribution over other statistics that strongly
weight high-density regions.

The more recently completed 1.2Jy \iras\ survey (Fisher 1992)
allows a more reliable determination of the \iras\ PDFs in a larger
volume and with more quantitative error analysis.
The moments of the density PDF from this survey indeed seem to behave as
expected from an initially Gaussian field subject to gravity
and $n\approx0$ near the smoothing scale, with
$s/\sigma=34/7-(n+3) \sim$
roughly constant as a function of smoothing scale, in the
range $1 - 2$ (Bouchet, Davis and Strauss 1992).
The skewness measured
from the QDOT study of counts in their 1-in-6 redshift survey of \iras\
galaxies (Saunders \etal 1991), given their errors,
is also consistent with the above measurements
(G. Efstathiou, private communication).

\bigskip
\cl{\bf 5. DISCUSSION AND CONCLUSION}

We investigated the quasilinear effects on the one-point probability
density functions (PDFs) of
initially Gaussian fluctuation fields. The laminar Zel'dovich
approximation provides useful analytic expressions that are confirmed and
extended into the multistreaming regime using a standard CDM N-body
simulation.
We found that the velocity PDF smoothed on scales $\geq 6\hmpc$ is hardly
affected by nonlinear evolutionary effects while the density PDF
develops a lognormal shape.

 The observed velocity and density fields, based on \pot\ reconstruction
from radial velocities with $\Omega=1$
or on an analysis of a redshift survey of \iras\
galaxies, have PDFs that are apparently consistent with Gaussian initial
conditions.  The
data used here, however, are still limited in volume and they still carry
large errors.  Noting that random errors can produce a spurious Gaussian
PDF, we were careful to estimate the uncertainties due to measurement
errors and to conservatively restrict ourselves to low-error regions at
the expense of sampling a larger volume.  New data
are expected to allow a significantly more accurate
determination of the PDFs.

 These results sound encouraging for the ``standard" model but can we
actually use them to reject any non-Gaussian model of interest?  Besides
the fact that the current errors are large, we are facing several
fundamental limitations.  For example, there is a general reason for the
velocity field to become Gaussian under a wide range of conditions, even
when it came from non-Gaussian initial fluctuations (Scherrer 1992).
Recall that the
peculiar velocity at a given point is related to the net peculiar force
there (directly proportional to it in the linear regime and in the
Zel'dovich approximation), which is the integral of the forces from all
the mass fluctuations around it.  Assume, for example, that the
non-Gaussianity is expressed as excessively large density peaks or wells
which dominate the large-scale force.  If the characteristic separation
between these structures is not larger than the typical range over which
the force converges, then the velocity is practically a sum over a few
independent random fluctuations and as such, based on the central limit
theorem, it becomes a Gaussian variable.  Thus, only certain non-Gaussian
models would show a non-Gaussian velocity PDF, so it has to be
individually evaluated for each model before the model can be rejected
based on an observed Gaussian velocity PDF.

 For instance, the Texture model cannot be rejected based on the
velocity PDF (Gooding \etal 1992). But
 it still remains to be seen whether the
model could be rejected based on the density PDF.
As another
example, it is clear that the non-Gaussianity scale in certain versions
of the cosmic string model is small enough for it not to have any
noticeable trace in the velocity PDF. A string model that could probably
be rejected is the version where the structure is dominated by $\sim
40\hmpc$ wakes formed behind long, well-separated strings which accrete
``hot" dark matter (e.g.  Brandenberger 1991).  In this model the
velocity at a point is typically determined by the nearest wake and is
therefore expected to be very non-Gaussian.

 The discriminatory power of the density PDF is also limited for several
reasons.  First, the deviation from Gaussianity depends on the \rms
fluctuation of the dynamical mass density, $\sigma$, which is deduced
from the observed $\sigma$ of galaxies (e.g. in the case of \iras) only
via a certain assumption concerning biasing --- the relation between the
galaxy density and the underlying mass density fluctuations which is not
well determined (see Dekel and Rees 1987 for a review).  Second, the
actual deviation from Gaussianity, and in particular its dependence on
the smoothing scale, is somewhat dependent on the shape of the power
spectrum of fluctuations.  We assumed above a ``standard" CDM spectrum
which seems to fit the data used here reasonably well, but recall that it
is not the Gaussian CDM model that one is trying to reject here.  The
moral is, again, that the density PDF has to be evaluated individually for
each specific non-Gaussian model.

 The main purpose of this paper was to provide useful tools for addressing
the question of Gaussian versus non-Gaussian initial fluctuations.  The
current preliminary comparison of observation with theory is encouraging
for the ``standard" model but it is certainly far from being conclusive.
We hope to be able to more quantitatively rule out certain non-Gaussian
models of interest with data that
are becoming available.

\bigskip
\cl{\bf ACKNOWLEDGEMENT}

We thank M. Davis, J. Huchra, M. Strauss and A. Yahil and our \pot\
colleagues for allowing the use of the 1.9Jy \iras\ and \pot\ data.
Supercomputer time was provided by the Cornell National Supercomputer
Facility.
This research has been supported by
US-Israel Binational Science Foundation grant 89-00194,
by NSF grant AST90-01762,
and by an Alfred P. Sloan Foundation Fellowship to E.B.
LK especially thanks the hospitality of the Hebrew University and MIT.

\page

\centerline{\bf APPENDIX A: Derivation of equation (25)}
\medskip

In this Appendix we derive equation (25) for the density PDF
 from the double  integral (22).
The main  work is to find the appropriate integration limits
in (22) in terms of $(\nu_1, \nu_2, \nu_3)$,  for which all three roots
$\lambda(\nu_1, \nu_2, \nu_3)$ of the cubic polynomial equation (24)
are real.

We  rewrite equation  (24)  in the canonical form:
$$
\lam^3 +A \lam^2 + B \lam +C=0.   \eqno(A1)
$$
First, let us  find the region of the $(A,B,C)$-space,
for which the three roots of equation (A1) are real.
To analyze the properties of the roots of this cubic equation, we
need its discriminant
$$
\Delta={1 \over 4} \biggl[ 2 \left({ A \over3}\right)^3
 -{A B \over 3} +C \biggr]^2
 -{1 \over 27} \left( {A^2 \over 3} -B \right)^3.   \eqno(A2)
$$
All three roots $\lam$ are real if this determinant is negative:
$$
\Delta \le 0.       \eqno(A3)
$$
To satisfy this condition, at least the second term on the right-hand side
of equation (A2) has to be negative, i.e.  $A^2-3B \ge 0$. Then we can write
the determinant in the form
$$
\Delta={1 \over 4}\,\left[C-C^{(-)}(A,B))\right]\left[C-C^{(+)}(A,B))\right],
\ 27 C^{(\pm)}(A,B)=-2A^3+9AB \pm 2 (A^2-3B)^{3/2}.\eqno(A4)
$$
The condition (A3) is satisfied if
$$
A^2\ge 3B\quad\hbox{and}\quad C^{(-)}(A,B) \le C \le C^{(+)}(A,B).    \eqno(A5)
$$
These are the conditions required for equation (A1) to have real roots.

Now we  find the corresponding region in the   $( \nu_1, \nu_2, \nu_3)$ space
where all three $\lam$ are real.  Comparing equations (24) and (A1), we have
$$
A=(\nu_1-\nu_2-1)D^{-1}\ ,\quad B=(\nu_2+\nu_3) D^{-2}\ ,\quad
C=-\nu_3 D^{-3}\ .\eqno(A6)
$$
Substituing equations (A6) into (A5),
we get two constraints in terms of   $( \nu_1, \nu_2, \nu_3)$:
$$
9(\nu_1-\nu_2+2)\nu_3+9(\nu_1-\nu_2-1)\nu_2-2(\nu_1-\nu_2-1)^3+
2\left[(\nu_1-\nu_2-1)^2-3(\nu_2+\nu_3)\right]^{3/2} \ge 0, \eqno(A7)
$$
and
$$
9(\nu_1-\nu_2+2)\nu_3+9(\nu_1-\nu_2-1)\nu_2-2(\nu_1-\nu_2-1)^3
-2\left[(\nu_1-\nu_2-1)^2-3(\nu_2+\nu_3)\right]^{3/2} \le 0. \eqno(A8)
$$
In addition, the expression in square brackets must be nonnegative.
We denote it henceforth as
$$
t^2=(\nu_1-\nu_2-1)^2-3\nu_2-3\nu_3, \eqno(A9).
$$
It is also convenient to introduce
$$
s=\nu_1-\nu_2+2. \eqno(A10)
$$

The constraints become simpler in terms of $\nu_1$ and the new variables
$s$ and $t$:
$$
|t|^3-{3 \over2}st^2+{1 \over2}(s^3-27\nu_1) \ge 0,~~
-|t|^3-{3 \over2}st^2+{1 \over2}(s^3-27\nu_1) \le0 . \eqno(A11)
$$
The intervals of $t$ which satisfy to both of these  constraints
simultaneously are
$$
t^2 >t_1^2(s,\nu_1)\quad\hbox{or}\quad
t_3^2(s,\nu_1)<t^2<t_2^2(s,\nu_1)\ ,  \eqno(A12)
$$
Here the functions $t_n(s,\nu_1)$ are defined by
$$
t_n(s,\nu_1)=s \left({1 \over2}+\cos\theta_n\right)\ ,\quad
\theta_n={1\over3}\arccos(54 s^{-3}\nu_1-1)\ .\eqno(A13)
$$
There are three roots because one may add an integer multiple of
$2\pi/3$ to $\theta$.  For definiteness, we label the roots according
to the phase $\theta$: $0\le\vert\theta_1\vert\le\pi/3$,
$\pi/3\le\vert\theta_2\vert\le2\pi/3$, $2\pi/3\le\vert\theta_3\vert\le\pi$,
so that $t_3^2\le t_2^2\le t_1^2$.
In addition to the constraints imposed on $t$ by equation (A12), we
require that $t_n(s,\nu_1)$ be real, implying
$$
s \ge 3 \nu_1^{1/3},~~ \nu_1 \ge 0, \eqno(A14)
$$
and
$$
s \le -3|\nu_1|^{1/3},~~ \nu_1 \le 0. \eqno(A15)
$$

The constraints (A12)--(A15)  give us the limits  in terms of $t$ and $s$.
They are easily expressed in terms of $\nu_2$ and $\nu_3$ using equations
(A9) and (A10) to get $\nu_2=\nu_1-s+2$ and
$$
\nu_3(t,s,\nu_1)={1\over3}\left(s^2-t^2\right)-s-\nu_1+1\ .\eqno(A16)
$$

Now we are ready to treat the integral (22).
We have from equations (19)--(21) (the Jacobian in eq. 21 simply removes
the three $\lambda$-dependent factors from eq. 19):
$$
\eqalign{ {\cal Q}_{\nu_1}&=\int d\nu_2 \int d\nu_3\,
{\cal Q_\nu}(\nu_1,\nu_2,\nu_3)\cr
&={5^{5/2}27 \over 8\pi(\sigma_{in}D)^6}\int d\nu_2 \int d\nu_3\,
\exp{\left[-{3(\nu_1-\nu_2-1)^2 \over (\sigma_{in}D)^2}+
{15\over2(\sigma_{in}D)^2}(\nu_2+\nu+3)\right]}\ .\cr}
                                                             \eqno(A17)
$$
The first integral over $\nu_3$ has limits given by equations (A12) and
(A16) and is trivial.  With two ranges of integration (one an unbounded
interval) this integral yields three exponentials.  The integral over
$\nu_2$ is performed using the variable $s$ through equation (A10).
Its limits are given by equations (A14) and (A15).  Their are two terms,
one for $\nu_1>0$ and the other for $\nu_1<0$ in equation (23).  The
integration of $s$ finally reduces to equation (25) of the main text.
\bigskip\bigskip

\centerline{\bf APPENDIX B: Asymptotics of the  density PDF for $\sigma \ll 1$}
\medskip
In this appendix we show that the density PDF given as the integral (25)
reduces to the Gaussian distribution (1) in the limit of
small density dispersion $\sigma \ll 1$.

To see this, we investigate the properties of the integral (25)
as $\sigma \ll 1$.
The expression (25) can be represented as an algebraic combination of
six integrals
$$
\Qs(\ti\rhos) = {N  \over \ti\rhos^2 \sigma^4}
\sum_{k=1}^{6} (\pm)I_k(p) \ ,\quad
I_k(p)=\int_{s_0}^{\infty} ds\,  e^{-p F_k(s)}. \eqno(B1)
$$
Here $p=1/\sigma^2 \gg 1$ is a large parameter and
$s_0=3\ti\rhos^{-1/3}$ denotes the lower limit of integration.
The six integrals come from the six terms in equation (25) that
one gets by multiplying out the exponentials.  Four of these terms
have a positive sign and two have a negative sign.
The formulas for $F_k(s)$ are rather tedious; some that we will use are
given by equations (B3) and (B4) below.

The asymptotic expansions of the integrals in equation (B1)
for $p \to \infty$ are
 $$
I_k(p) \sim e^{-p F_k(s_0)} \cdot \hbox{(asymptotic series in powers
of $p^{-1}$)}\ . \eqno(B2)
$$
The asymptotic series have to be defined
based on the analytic properties of the functions $F_k(s)$.
Using equation (25b) for the functions $\beta_n(s)$, after straightforward
but  tedious analysis of the functions $F_k(s)$ in the vicinity of $s_0$,
one can show that two following terms in the sum (B1) are of leading
order as $p \to \infty$: the positive term we denote $I_2$ with
$$
F_2(s)=(s-3)^2/2+\beta_2^2(s)/2,  \eqno(B3)
$$
and the negative term we denote $I_3$ with
$$
F_3(s)=(s-3)^2/2+\beta_3^2(s)/2.  \eqno(B4)
$$
The other integrals $I_1$, $I_4$, $I_5$, and $I_6$ are exponentially
suppressed compared with $I_2$ and $I_3$.  Thus we have to consider the
combination $(I_2-I_3)$ for $p \to \infty$.
The decompositions of the functions $F_{2,3}$ in the vicinity of $s_0$ are:
$$
F_2(s)={(s_0 -3)^2 \over2}
 +\biggl({{7s_0-3} \over 2} \biggr)\cdot(s -s_0)-
 { {5\sqrt{s_0}} \over 3} \cdot (s-s_0)^{3/2}
-{1 \over 3}\cdot (s -s_0)^2 + ...,  \eqno(B5)
$$
$$
F_3(s)={(s_0 -3)^2 \over2}
 +\biggl({{7s_0-3} \over 2} \biggr)\cdot(s -s_0)+
 { {5\sqrt{s_0}} \over 3} \cdot (s-s_0)^{3/2}
-{1 \over 3}\cdot (s -s_0)^2 + ... .  \eqno(B6)
$$

To obtain an asymptotic series (B2), the functions $F_k$ involved in
the integrals $I_k$, have to be represented as analytic functions of the
variable of integration. However, the series given by (B5) and (B6) show that
$F_2(s)$ and $F_3(s)$ are not analytic functions of $s$.  Therefore we must
change variables of integration.  For this purpose we introduce
$$
v=\sqrt{s-s_0}.     \eqno(B7)
$$
Now the integrals  read as
$$
I_k(p)=2 \int_0^{\infty} dv v e^{-p {\tilde F_k(v)}}, \eqno(B8)
$$
where the functions of the new variable have the following analytic series in
the vicinity of $v=0$:
$$
\tilde F_{2,3}(v)={(s_0 -3)^2 \over2}
 +\biggl({{7s_0-3} \over 2} \biggr)\cdot v^2 \pm
 { {5\sqrt{s_0}} \over 3} \cdot v^3
-{1 \over 3}\cdot v^4 + ... , \eqno(B9)
$$
The upper sign corresponds to $k=3$ and the lower to $k=2$.
For these analytic functions  we can use the general
formula of the asyptopic expansion (e.g. Olver 1974)
$$
I_k(p) \simeq  2 e^{-p{\tilde F_k(0)}} \sum_{n=0}^{\infty}
\Gamma \left({{n+2} \over 2}\right){{a_n^{(k)}} \over {p^{(n+2)/2}}},
\eqno(B10)
$$
where $\Gamma(x)$ is the Gamma function.
The coefficients $a_n^{(2,3)}$ are expressed through the coefficients of the
series (B9). The first two of them, which we will use, are
$$
 a_0^{(2,3)}=\left( 7s_0-6 \right)^{-1}, \eqno(B11a)
$$
and
$$
a_1^{(2,3)}=(\pm 1){{-\sqrt{10s_0}}\over{(7s_0-6)^{5/2}}}.\eqno(B11b)
$$

In the sum (B1) we have the combination $(I_2-I_3)$. Substituing in this
 combination
the asymptotic series (B10) and using (B11),  we find that the first
terms with coefficients $a_0$ cancel.  The leading remaining terms with
coefficients $a_1^{(2,3)}$ give
$$
I_2-I_3={{2\sqrt{10\pi s_0}}\over{(7s_0-6)^{5/2}}}\sigma^3
 e^{-(s_0-3)^2 /2\sigma^2}, \eqno(B12)
$$
where we have replaced $p$ by the original the small parameter $\sigma$.

Now we are ready to reproduce the result of the linear theory.
Let us recall that $s_0={3 \tilde\rhos^{-1/3}} \approx 3-D(t)\delta$
for small density fluctuations $\delta \ll 1$ and
$\sigma=D(t)\sigma_{in}$. Substituing these formulas into (B12), we obtain
$$
I_2-I_3=\left( 9 \cdot 5^{3/2}/4\pi \right)^{-1}
 {1 \over {\sqrt{2\pi}\sigma_{in}}}e^{-\delta^2 /2\sigma_{in}}.
\eqno(B13)
$$
Substituing this expression into equation (B1) we get the normal distribution
for the density fluctuations $\delta$.

\page
\cl{\bf REFERENCES}
\def\ref{\par\noindent\hangindent 15pt}
\def\apj#1{{ ApJ} { #1}}
\def\apjl#1{{ ApJ} { #1}}
\def\mn#1{{ M.N.R.A.S.} { #1}}
\def\aa#1{{ A\&A} { #1}}
\def\nat#1{{ Nature} { #1}}

\ref Bahcall, N. 1988, { Ann. Rev. Astr. Ap.} { 26}, 631.
\ref Bertschinger, E. 1989, { Ann. Ny. Acad. Sci.} { 57}, 151.
\ref Bertschinger, E. and Dekel, A. 1989, \apjl{336}, L5.
\ref Bertschinger, E., Dekel, A., Faber, S.M., Dressler, A. and Burstein, D.
    1990, \apj{364}, 370.
\ref Bertschinger, E. and Gelb, G. 1991, { Computers in Physics} Mar/Apr,
     164.
\ref Bertschinger, E., Gorski, K. and Dekel, A. 1990, \nat{345}, 507.
\ref Bouchet, F. R., Davis, M. and Strauss, M. 1992, in: {Proc. of the 2nd DAEC
     Meeting}, ed. Mamon, G. and Gerbal, D., (Paris Observatory), p.287.
\ref Weinberg, D. and Cole, S. 1992, \mn{259}, 652.
\ref Brandenberger, R. 1991, Physica Scripta, { T36}, 114.
\ref Coles, P. and Frenk, C. S. 1991, \mn{253}, 727.
\ref Coles, P. and Jones, B. 1991, \mn{248}, 1.
\ref Davis, M., Efstathiou, G., Frenk, C. and White, S.D.M. 1985,
       \apj{292}, 371.
\ref Dekel, A. 1981, \aa{101}, 79.
\ref Dekel, A., Bertschinger, E. and Faber, S.M. 1990, \apj{364}, 349.
\ref Dekel, A., Bertschinger, E., Yahil, A., Strauss, M. Davis, M.
     and Huchra, J. 1993, \apj{}, in press.
\ref Dekel, A. and Rees, M.J. 1987, \nat{326}, 455.
\ref Doroshkevich, A.G. 1970,{ Astrofizica}, { 6}, 581
\ref Efstathiou, G. 1990, in: { Physics of the Early
     Universe},  eds. J. Peacock, A. Heavens, and A. Davies,
     (Inst. of Physics Publishing, Bristol), p.361.
\ref Efstathiou, G., Kaiser, N., Saunders, W., Lawrence, A.,
     Rowan-Robinson M., Ellis, R.S., and Frenk, C.S. 1990, \mn{247}, 10p.
\ref Fisher, K.B. 1992, PhD. thesis, University of California, Berkeley
\ref Gooding, A., Park, C., Spergel, D., Turok, N. and Gott, J.R.,III.
     1992, \apj{393}, 42.
\ref Hamilton, A. 1985, ApJ,{ 292}, L35.
\ref Juszkiewicz, R., Bouchet, F. R. and Colombi, S. 1993, Ap.J., submitted.
\ref Kofman, L. 1991a, Physica Scripta, { T36}, 108.
\ref Kofman, L. 1991b, in:{ Primordial Nucleosynthesis and Evolution of
     Early Universe}, eds. Sato, K. and Audouze, J. (Dordrecht: Kluwer), p.495.
\ref Kofman, L., Melott, A. Pogosyan, D. Yu. and Shandarin , S. 1992,
     \apj{393}, 437.
\ref Kolatt, T., Seljak, U., Dekel, A. and Bertschinger, E. 1993,
     in preparation.
\ref Linde, A. 1990, { Particle Physics and Inflationary Cosmology}
    (Harwood, Chur, Switzerland).
\ref Lynden-Bell, D., Faber, S.M., Burstein, D., Davies, R.L., Dressler,
     A., Terlevich, R.J. and Wegner, G. 1988, \apj{326}, 19.
\ref Maddox, S.J, Efstathiou, G., Sutherland, W.J. and Loveday, J.
     1990, \mn{242}, 43p.
\ref Melott, A. L. and Shandarin, S. F. 1989, \apj{343}, 26.
\ref Monin, A. and Yaglom, A., 1971, { Statistical fluid mechanics}
     (Camdridge: MIT Press).
\ref Nityananda, R. and Ostriker, J. 1984, J. Astr. Ap. { 5}, 235.
\ref Nusser, A., Dekel, A., Bertschinger, E. and Blumenthal, G. R. 1991,
     \apj{379}, 6.
\ref Olivier et al. 1993, preprint.
\ref Olver, F.W.J. 1974, { Introduction to Asymptotics and Special
      Functions} (NY: Academic Press).
\ref Ostriker, J. P. 1988, in: { IAU Symp. No. 130}, ed. J. Audouze and A.
     Szalay (Dordrecht: Reidel).
\ref Peebles, P.J.E. 1980, { The Large-Scale
     Structure of the Universe} (Princeton: Princeton University Press).
\ref Rowan-Robinson, M., Lawrence, A., Saunders, W., Crawford, J.,
     Ellis, R.S., Frenk, C.S., Parry, I., Xiaoyang, X.,
     Allington-Smith, J., Efstathiou, G. and  Kaiser, N.
     1990, \mn{247}, 1.
\ref Saunders, W., Frenk, C.S., Rowan-Robinson, M., Ellis, R., Lawrence, A.,
     Kaiser, N., Efstathiou, G., Crawford, J., Xia, X.Y., and Parry, I. 1991,
     \nat{349}, 32.
\ref Scherrer, R. \& Bertschinger, E. 1991, \apj{381}, 349.
\ref Scherrer, R. 1992, \apj{390}, 330.
\ref Shandarin, S.F. and Zel'dovich, Ya.B. 1989,
     { Rev.Mod.Phys.} { 61}, 185.
\ref Strauss, M. A., Davis, M., Yahil, A., and Huchra, J. P. 1990,
     \apj{361}, 49.
\ref Strauss, M. A., Davis, M., Yahil, A., and Huchra, J. P. 1992,
     \apj{385}, 421 .
\ref Trimble, V. 1987, { ARAA}, 25, 425.
\ref Turok, N. 1991, { Physica Scripta}, T36, 135.
\ref Yahil, A., Strauss, M. A., Davis, M., and Huchra, J. P. 1991,
     \apj{372}, 380.
\ref Zel'dovich, Ya.B. 1970, \aa{5}, 20.

\page
\cl{\bf FIGURE CAPTIONS}
\def\fig#1{\noindent\hangindent 15pt {\bf Figure #1 :\ }}

\fig{1} The mean number of streams at a point according to the Zel'dovich
approximation, $N_s$,
as functions of the linear standard deviation of density fluctuations,
$\sigma_\delta = a(t) \sigma_{in}$.

\fig{2} The projected distribution of matter in a slice of the
standard CDM N-body simulation used in this paper
at time corresponding to linear $\sigma_8=1$.
The box side is $200\hmpc$ and the slice thickness is $25\hmpc$.

\fig{3} Velocity PDF $P(v/\sigma_v)$ in the N-body simulation
 (at $a=\sigma_8=1$).
The distribution has been assembled from $80^3$ cubic grid points inside
the box of side $200\hmpc$.
The error bars are the standard deviations of $P$ in the eight
octants of the volume.
Shown is the most nonlinear case, $a=1$ and $R_s=6\hmpc$,
in comparison with a Gaussian distribution.

\fig{4} Density PDF $P(\rho/\bar\rho)$ in the
N-body simulation (as in the previous figure, $a=1$)
for three different smoothing scales: $R_s=18, 12, 6\hmpc$,
corresponding to $\sigma=0.11, 0.26, 0.55$.
The solid curves are lognormal distributions with the corresponding
$\sigma$ and the dashed curves are laminar Zel'dovich
distributions with the same $\sigma$.

\fig{5} Moments of the distribution of density and velocity in the
N-body simulation. Shown are the skewness and kurtosis as functions of
the standard deviation.
Each curve corresponds to a given time in the simulation, $a=0.5, 0.7,
1$, and the variable along each curve is the Gaussian smoothing radius
in the comoving range $6\leq R_s\leq 21\hmpc$.
The error bars are $\pm$ standard deviation in the eight
octants of the volume.
The dotted line marks the second-order prediction (Peebles 1980)
$s_\delta=(34/7)\sigma_\delta$.

\fig{6} PDFs for \pot\ velocity and density fields ($\Omega=1$)
with $R_s=12\hmpc$ smoothing
(Dekel \etal 1990; Bertschinger \etal 1990)
in a volume (selected to have small errors)
comparable to a sphere of radius $37\hmpc$.
Also shown are the Gaussian and lognormal curves with the same
$\sigma$.
The error bars correspond to distance measurement errors as derived by
Monte Carlo simulations.
The relative errors associated with the limited volume sampled are
roughly 57\% larger than the errors shown in the next figure
based on the N-body simulation.

\fig{7} Mean and standard deviation of the PDFs
in eight disjoint spheres of radius $50\hmpc$ in the
N-body simulation with smoothing $R_s=12\hmpc$.

\fig{8} PDFs for \iras\ 1.9Jy density and velocity fields
(Strauss \etal 1991, Yahil \etal 1991) in a sphere of radius $80\hmpc$.
(a) $R_s=12\hmpc$ and (b) $R_s=6\hmpc$.
Also shown are the Gaussian and lognormal curves with the same
$\sigma$.
The errors associated with the limited volume sampled should be roughly
twice the errors shown in figures 3 and 4 based on the whole N-body
simulation, or half the errors in figure 7.

\bye